\documentstyle[prd,aps,epsf]{revtex}

\begin{document}
\draft
\title{Testing models with a nonminimal Higgs sector through the decay $t\to q+WZ$}
\author{J. L. D\'{\i}az Cruz and D. A. L\'opez Falc\'on}
\address{Instituto de F\'{\i}sica, BUAP, Apartado Postal J-48, 72570 Puebla, Puebla, M\'exico.}
\date{Received 9 November 1999}
\maketitle

\begin{abstract}
We study the contribution of the charged Higgs boson to the rare decay of the top quark $t\to q+WZ$
($q=d,s,b$) in models with Higgs sectors that include doublets and triplets. Higgs doublets are needed
to couple a charged Higgs with quarks, whereas the Higgs triplets are required to generate the
nonstandard vertex $HWZ$ at the tree level. It is found that within a model that respects the
custodial $SU(2)_c$ symmetry and avoids flavor--changing neutral current (FCNC) by imposing discrete
symmetries, the decay mode $t\to b+WZ$ can reach a branching ratio (BR) of order $10^{-2}$, whereas
the decay modes $t\to (d,\,s)+WZ$, can reach a similar BR in models where FCNC are suppressed by
flavor symmetries.
\end{abstract}

\section*{INTRODUCTION}
The mass of the top quark, which is larger than any other fermion mass in the standard model (SM) and
almost as large as its scale of electroweak symmetry breaking (EWSB), cannot be explained within the
SM~\cite{smrev}. This has originated speculations about the possible relationship between the top
quark and the nature of the mechanism responsible for EWSB. Several models have been proposed, where
such large mass can be accommodated or plays a significant role. In the supersymmetric (SUSY)
extensions of the SM~\cite{susyrev}, the large value of the top quark mass can drive the radiative
breaking of the electroweak symmetry; furthermore within the context of SUSY grand unified theories
(GUT's) the fermions of the third--family can be accommodated in scheemes where their masses arise
from a single Yukawa term~\cite{gutyuka}. On the other hand, in some top--condensate (TC)
models~\cite{topcmod} it is postulated that new strong interactions bind the heavy top quark into a
composite Higgs scenario.

From a more phenomenological point of view, it is also intriguing to notice that the top quark decay
seems to be dominated by the SM mode ($t\to bW$), not only within the SM but also in theories beyond
it, which makes the top quark decay width almost insensitive to the presence of new physics; unless
the scale of new physics is lighter than the top quark mass itself, such that new states can appear in
its decays. This is the case, for instance, in the general two Higgs doublet model
(THDM--III)~\cite{modelIII}, where the flavor changing mode $t\to c+h$ can be important for a light
Higgs boson ($h$), or in SUSY models with light stop quark and neutralinos, in whose case the decay
$t\to\tilde{t}+\tilde{\chi^0}$ can also be relevant. But in general, the rare decays of the top quark
have undetectable branching ratios (BR's); for instance, the flavor--changing neutral current (FCNC)
rare decays $t\to cV$ ($V=\gamma,Z,g$) have a very small BR in the SM, of the order
$10^{-11}$~\cite{ourfcnc}, and are out of reach of present and future colliders. A similar result is
obtained in several extensions of the SM; for instance in the THDM--II, minimal SUSY extensions of the
SM (MSSM) and left--right models, to mention some cases~\cite{tcV}.

The rare decay $t\to qWZ$ $(q=b,s,d)$, may be above the threshold for the production of a real $WZ$
state, provided that $m_t\geq m_q+m_W+m_Z$. The possibilities to satisfy this relation depend on the
final state $q$ and the precise value of the top quark mass, which according to the Particle Data
Group~\cite{PDG98} is $m_t=173.8\pm 5.2\,$~GeV. For the case when $q=b$, the top quark mass must
satisfy $m_t \geq 176.1\pm 0.5$~GeV, where the uncertainty on the right--hand side is mostly due to
the ambiguity in the bottom quark mass, thus $t\to bWZ$ can occur on--shell only if $m_t$ takes its
upper allowed value (at $1\sigma$). However, if $q=d$ or $s$, the decays $t\to qWZ$ can occur even
when $m_t$ takes its central value.

The value of BR($t\to bWZ$) predicted in SM is $5.4\times 10^{-7}$~\cite{tbWZinSM}, which is beyond
the sensitivity of Tevatron Run II or even CERN Large Hadron Collider (LHC); thus its observation
would truly imply the presence of new physics. For $q=d,s$ the SM result is even smaller, since the
amplitude is supressed by the Cabibbo--Kobayashi--Maskawa (CKM) matrix elements $V_{tq}$. On the other
hand, the decay mode $t\to qWZ$ can proceed through an intermediate charged Higgs boson that couples
to both $tq$ and $WZ$ currents, and thus can be used to test the couplings of Higgs sectors beyond the
SM~\cite{phenoextSM}.

The construction of extensions of the SM Higgs sector must satisfy the constraints impossed by the
successful phenomenological relation $\rho\equiv m_W^2/m_Z^2\cos\theta_w=1$, which also measures the
ratio between the neutral and charged current couplings strength. At tree level this relation is
satisfied naturally in models that include only Higgs doublets, but in more general scenarios, there
could be tree level contributions to $\rho-1$. Since the vertex $HWZ$ arises at tree level only for
Higgs bosons lying in representations higher than the usual SM Higgs doublet, there could be
violations of the constraints impossed by the $\rho$ parameter. However, tree--level deviations of the
electroweak $\rho$ parameter from unity can be avoided by arranging the nondoublet fields and the
vacuum expectation values (V.E.V's) of their neutral members, so that a custodial $SU(2)_c$ symmetry
is maintained~\cite{GVW}. On the other hand, a generic coupling of the charged Higgs boson with
fermions may be associated with the possible appearence of FCNC in the Higgs--Yukawa sector. FCNC are
automatically absent in the minimal SM with one Higgs doublet, however in multiscalar models large
FCNC can appear if each quark flavor couples to more than one Higgs doublet~\cite{HiggsHG90}. FCNC can
be avoided either by impossing some {\it ad hoc} discrete symmetry to the Yukawa Lagrangian, i.e., by
coupling each type of fermion only to one Higgs doublet; or by using flavor symmetries. The former
case is used in the so--called two Higgs doublet models I and II, whereas the last one is associated
with model III, here FCNC is only suppressed by some ansatz for the Yukawa matrices, for instance the
Li--Sher one: $(Y_q)_{ij}=\sqrt{m_i m_j}/m_W$, whose phenomenology was studied
in~\cite{modelIIIpheno,ourtcvv}.

In this Rapid Communication we shall consider, in a very general setting, the contribution of a
charged Higgs boson to the decay $t\to qWZ$, and present the results in terms of two factors that
parametrize the doublet--triplet mixing and the nonminimal Yukawa couplings, respectively. Then, we
discuss the values that these parameters can take for specific extensions of the SM, when the
constraints from both the custodial symmetry and FCNC are satisfied, and present the predicted values
for BR($t\to qWZ$).
\vspace{15pt}

\section*{The decay $t\to qWZ$}
We are interested in studying the contribution of charged Higgs boson to the rare decay of the top
quark $t\to qWZ$ ($q=d,s,b$), within the context of models with extended Higgs sector that include
additional Higgs doublets and triplets. The charged Higgs will be assumed to be the lightest charged
mass eigenstate that results from the general mixing of doublets and triplets in the charged sector.
\footnote{This is justified by our explicit analysis of the Higgs potential for several models with
Higgs doublets and triplets, which will be presented elsewhere.}

Higgs doublets are needed in order to couple the charged Higgs with quarks; the vertex $tqH^\pm$ will
be written as follows
\begin{equation}
		\frac{ig}{2\sqrt{2}m_W}\eta_{tq}\cos\alpha
		\left[(m_t\cot\beta+m_q\tan\beta)+(m_t\cot\beta-m_q\tan\beta)\gamma_5\right],
\end{equation}
which can be considered as a modification of the result obtained for the Yukawa sector of the general
THDM, where $\cos\alpha$ is included to account for the doblet--triplet mixing; $\tan\beta$ is the
ratio of the V.E.V.'s of the two scalar doublets. The charged Higgs coupling to the quarks is also
determined by the parameters $\eta_{tq}$, which is equal to the CKM mixing matrix only for model--II,
i.e., $\eta_{tq}^{II}=V_{tq}$; however, in the general case (THDM--III), one can have
$\eta_{tq}^{III}>V_{tq}$~\cite{modelIII}.

On the other hand, we require a representation higher than the doublet, in order to obtain a sizeable
coupling $HWZ$ at tree level, which is written as
\begin{equation}
		-\frac{igm_W}{\cos\theta_w}\sin\alpha g_{\mu\nu}.
\end{equation}

In order to evaluate the decay $t\to qWZ$, we shall write a general amplitude to describe the
contribution of the intermediate charged Higgs, neglecting the SM contribution, which is a good
approximation since the corresponding BR is very suppressed. To calculate the amplitude one also needs
to take into account the finite width of the intermediate charged Higgs boson with momentum $p_H$,
mass $m_H$, and width $\Gamma_H$, for this we shall use the relativistic Breit--Wigner form of the
propagator in the unitary gauge. Then, the amplitude can be written in general as
\begin{equation}
		{\cal M}=A\left[\overline{u}(p_q)(a+b\gamma_5)u(p_t)\right]
			\left(\frac{-i}{p_H^2-\hat{m}_H^2}\right)
			\left[g_{\mu\nu}\epsilon_W^{*\mu}\epsilon_Z^{*\nu}\right],
\end{equation}
where $\hat{m}_H\equiv m_H+(i/2)\Gamma_H$; $a$, $b$, and $A$ are constants related to the parameters
$\alpha$ and $\eta_{tq}$ previously mentioned
\begin{eqnarray}
		a&=&m_t\cot\beta+m_q\tan\beta,\nonumber\\
		b&=&m_t\cot\beta-m_q\tan\beta,\\
		A&=&\frac{g^2}{2\sqrt{2}\cos\theta_w}\eta_{tq}\cos\alpha\sin\alpha.\nonumber
\end{eqnarray}

To calculate the partial decay width, we shall perform a numerical integration of the expression for
the squared amplitude, over the standard three--body phase space, namely
\begin{equation}
		\Gamma(t\to qWZ)=\frac{1}{(2\pi)^3}\frac{1}{32m_t^3}
		\int\mid\overline{{\cal M}}\mid^2ds\,dt.
\label{Width}
\end{equation}
$\mid\overline{{\cal M}}\mid^2$ denotes the squared amplitude, averaged over initial spins and summed
over final polarizations, it has the form
\begin{equation}
		\mid\overline{{\cal M}}\mid^2=\frac{\mid A\mid^2
		[(a^2+b^2)(m_t^2+m_q^2-s)+(a^2-b^2)2m_tm_q]}
		{(s-m_H^2)^2+m_H^2\Gamma_H^2}
		\left[2+\left(\frac{s-m_W^2-m_Z^2}{2m_Wm_Z}\right)^2\right].
\end{equation}
The integration limits are
\begin{equation}
			(m_W+m_Z)^2\leq s\leq (m_t-m_q)^2,
\end{equation}
and
\begin{equation}
				t^- \leq t\leq t^+,
\end{equation}
where
\begin{equation}
		t^\pm=m_t^2+m_Z^2-\frac{1}{2s}[(s+m_t^2-m_q^2)(s+m_Z^2-m_W^2)
		\mp\lambda^{1/2}(s,m_t^2,m_q^2)\lambda^{1/2}(s,m_Z^2,m_W^2)],
\end{equation}
and $\lambda(x,y,z)=(x+y-z)^2-4xy$.

The branching ratio for this decay is obtained as the ratio of Eq.~(\ref{Width}) to the total width of
the top quark, which will include the modes $t\to qW$ and $t\to qH$; the expressions for the widths
are
\begin{equation}
		\Gamma(t\to qW^+)=\frac{G_Fm_t^3}{8\pi\sqrt{2}}V_{tq}
		\left(1-\frac{m_W^2}{m_t^2}\right)^2\left(1+2\frac{m_W^2}{m_t^2}\right)
		\left[1-\frac{2\alpha_s}{3\pi}\left(\frac{2\pi^2}{3}-\frac{5}{2}\right)\right],
\end{equation}
and
\begin{eqnarray}
		\Gamma(t\to qH^+)&=&\frac{g^2}{128\pi m_W^2m_t}\eta_{tq}\cos^2\alpha
		\left[a^2[(m_t+m_q)^2-m_H^2]+b^2[(m_t-m_q)^2-m_H^2]\right]\nonumber\\
		&\times&\lambda^{1/2}\left(1,\frac{m_q^2}{m_t^2},\frac{m_H^2}{m_t^2}\right).
\end{eqnarray}
On the other hand, the Higgs width will include the fermionic decays into $H\to c\overline{s}$ and
$H\to\tau\nu_{\tau}$; adding them we obtain
\begin{equation}
		\Gamma(H^+\to f\overline{f'})=\frac{g^2m_H}{32\pi m_W^2}\cos^2\alpha
		\left[3\eta_{cs}(m_c^2\cot^2\beta+m_s^2\tan^2\beta)+m_\tau^2\tan^2\beta\right]
\end{equation}
as well as the bosonic mode $H\to WZ$
\begin{eqnarray}
	\Gamma(H^+\to W^+Z)&=&\frac{g^2m_H}{64\pi}\sin^2\alpha
		\left[	1+\left(\frac{m_W^2}{m_H^2}\right)^2+\left(\frac{m_Z^2}{m_H^2}\right)^2
	-2\frac{m_W^2}{m_H^2}-2\frac{m_Z^2}{m_H^2}+10\frac{m_W^2}{m_H^2}\frac{m_Z^2}{m_H^2}
		\right]\nonumber\\
	&\times&\lambda^{1/2}\left(\frac{m_H^2}{m_W^2},\frac{1}{\cos^2\theta_w},1\right).
\end{eqnarray}

\section*{Results and conclusions}
\begin{center}
\begin{figure}
\caption{Branching ratio for the decay $t\to bWZ$, as a function of $m_H$ for
$\cos\alpha\sin\alpha=\frac{1}{2}$ (upper curves) and $\sin\alpha=0.04$ (lower curves); and for
$\tan\beta=\frac{m_t}{m_b}$ (solid curves) and $\tan\beta=2$ (dashed curves); it is also assumed that
$\eta_{tq}=V_{tq}$. The line indicate the SM prediction.}
\label{BRtbWZ}
\vspace{12pt}
\caption{Branching ratio for the decay $t\to sWZ$, as a function of $m_H$ for
$\cos\alpha\sin\alpha=\frac{1}{2}$ (upper dashed and solid curve) and $\sin\alpha=0.04$
(middle curve); and for $\tan\beta=\frac{m_t}{m_b}$ (solid curve) and $\tan\beta=2$ (dashed curves);
taking also $\eta_{tq}=1$.}
\label{BRtsWZ}
\epsfysize=318pt \epsfbox{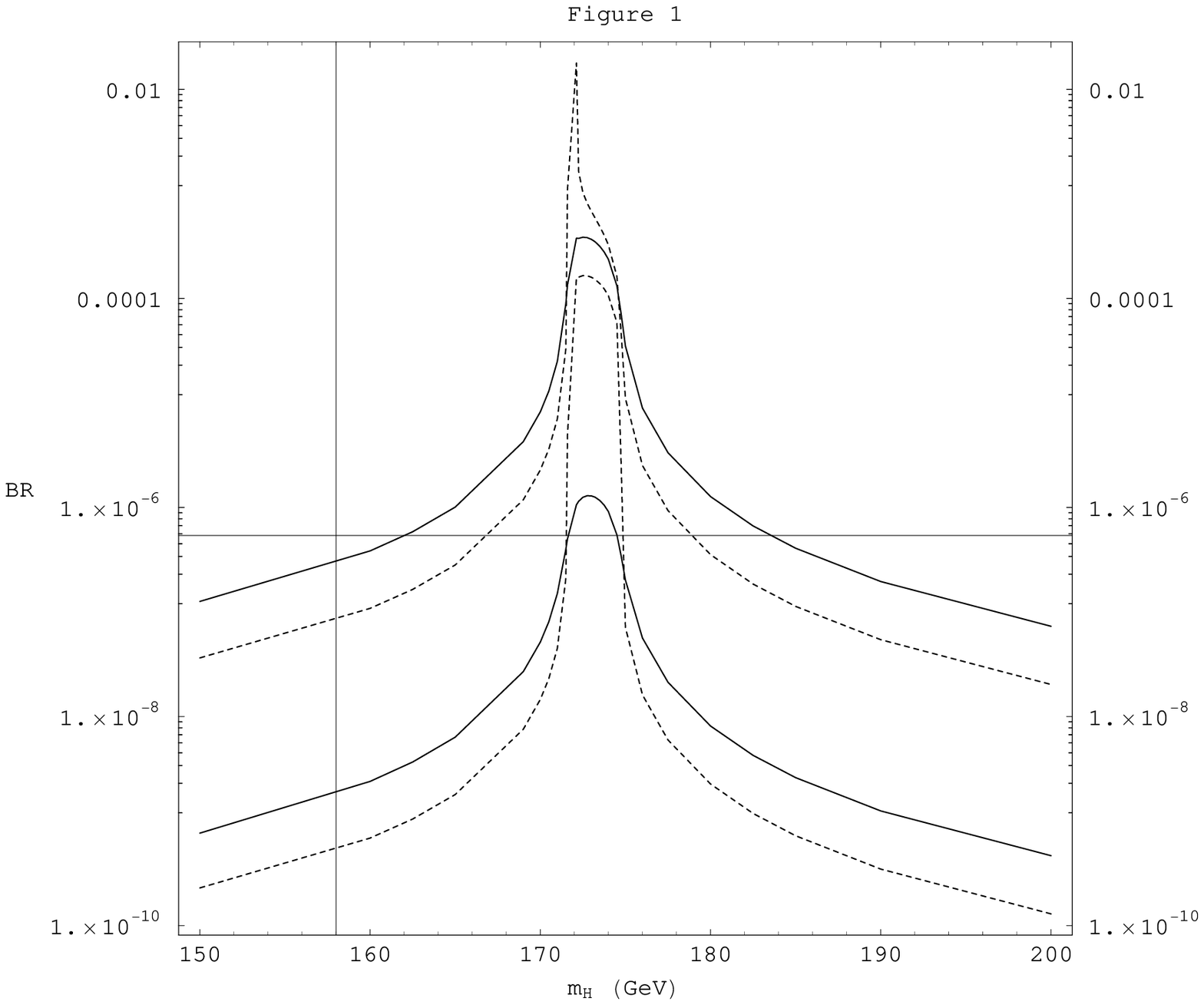}
\epsfysize=318pt \epsfbox{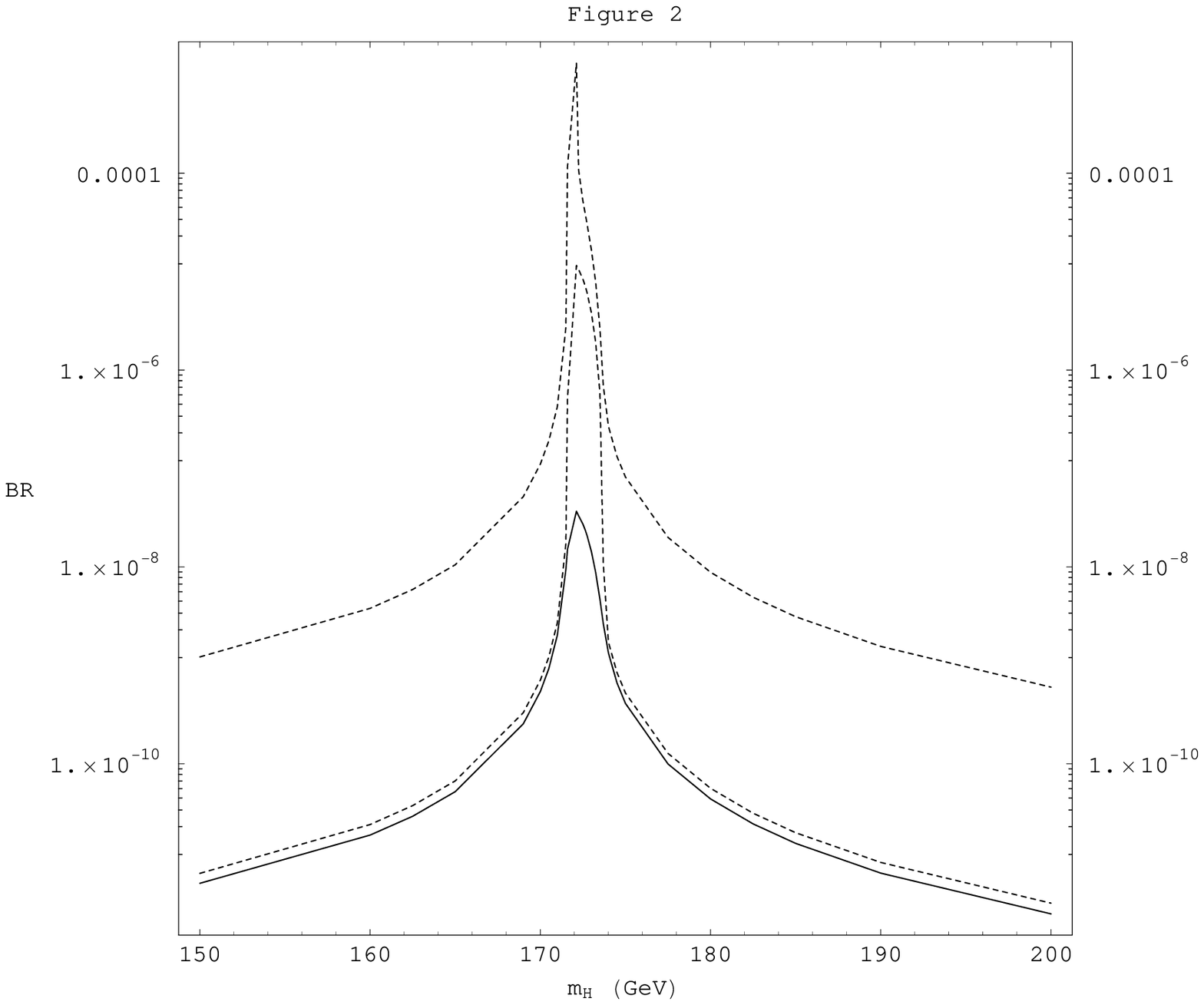}
\end{figure}
\end{center}

In order to present the results for the mode $t\to bWZ$, i.e., $q=b$, we shall assume that the top
quark mass takes its upper allowed value, and will consider a Yukawa sector similar to the model--II,
in whose case the factor $\eta_{tb}$ is equal to the CKM matrix element $V_{tb}$ $(\simeq 1)$; the
results are shown for two values of $\tan\beta$ (2, and $m_t/m_b$) which are acceptable for
GUT--Yukawa unification. For the factor $\cos\alpha\sin\alpha$, which is part of the constant $A$, we
shall consider first the value $\frac{1}{2}$, which corresponds to the maximum value that can be
expected to arise in an scenario where the custodial symmetry is respected, for instance in a model
with one Higgs doublet and two Higgs triplets of hypercharges 0 and 2, respectively, where one can
align the V.E.V.'s to respect the custodial symmetry and obtaining $\rho=1$\cite{GVW}.
\footnote{Although our framework is similar to the one of Ref.\ \cite{GVW}, in our case we are
allowing full mixing between all the scalar multiplets of the model, which allows us to have charged
and neutral Higgs bosons that couple simultaneously to both fermion and gauge boson pairs.} On the
other hand, to consider a model without a custodial symmetry, we take the value $\sin\alpha=0.04$,
which corresponds to the maximum value that is allowed by the experimental error in the $\rho$
parameter~\cite{rhostriction}. With all these considerations, we shown in Fig.\ \ref{BRtbWZ} our
results for the BR of the decay $t\to bWZ$; we notice that it can reach a maximum value of order
$1.78\times 10^{-2}$.

For the decays into the light quarks, still working within the framework of model II, we obtain a very
suppressed result, where we are taking now the central value for the top quark mass, namely, for
$t\to sWZ$ we get a maximum value for the BR of order $1.95\times 10^{-6}$ for
$\sin\alpha\cos\alpha=\frac{1}{2}$ and $\tan\beta=2$; for $t\to dWZ$ we get results even smaller and
thus uninteresting. On the other hand, if we consider a model with a Yukawa sector of the type
THDM--III, the coupling of the charged Higgs with the quarks is not determined by the CKM mixing
matrix, then the couplings $t\bar{d}H^-$ and $t\bar{s}H^-$ may not be suppresed. Although in model III
there can be dangerous FCNC, it happens that such effects have not been tested by top quark decays,
and thus can give large and detectable effects~\cite{ourtcvv}. For the parameter $\alpha$ we take the
same values of the previous case, assuming also $\eta_{ts}=\eta_{td}$ we get a maximum value for the
BR of order $1.31\times 10^{-3}$ for both $t\to (d,\,s)+WZ$, as shown in Fig.\ \ref{BRtsWZ}.

We conclude from our results that there exist a region of parameters where it is possible to obtain a
large BR for the decay $t\to bWZ$. Moreover, for $m_H=162$ to $m_H=182$~GeV,
$\cos\alpha\sin\alpha=\frac{1}{2}$, and $\tan\beta=m_t/m_b$ we obtain a BR larger than the one
predicted by the SM. Furthermore, the maximum value for the BR, of order $10^{-2}$, seems factible to
be detected at the future CERN LHC, where about $10^8$ top quark pairs could be produced, and one
would have $10^6$ events of interest with only one top quark decaying rarely. If we also include the
decays of the W and Z into leptonic modes, to allow a clear signal, one would end with about
$1.3\times 10^4$ events, which is interesting enough to perform a future detailed study of
backgrounds; however this is beyond the scope of present work. On the other hand, we observe from
Fig.\ \ref{BRtsWZ} that even within models without a custodial symmetry with $\sin\alpha=0.04$, it is
possible to get BR for the decay $t\to sWZ$ larger than the SM result, or the result obtained within
models where $\sin\alpha\cos\alpha=\frac{1}{2}$, depending on the value of $\tan\beta$; in some cases
it can reach a BR of order $1.31\times 10^{-3}$.

In conclusion, we find that the decay $t\to qWZ$ is sensitive to the contribution of new physics, in
particular from a charged Higgs boson, which makes this mode an interesting arena for testing physics
beyond the SM.

\acknowledgments
This research was supported in part by the Benem\'erita Universidad Aut\'onoma de Puebla with funds
granted by the Vicerrectoria de Investigaci\'on y Estudios de Posgrado under contract VIEP/930/99, and
in part by the CONACyT under Contract G 28102 E.

\twocolumn

\begin{references}

\bibitem{smrev} W. Hollik, Radiative Corrections: Applications of Quantum Field Theory to
Phenomenology, Proceedings Edited by Joan S\`ola, Singapore, World Scientific, (1999) p. 44.
\bibitem{susyrev} H. P. Nilles, Phys. Rep. {\bf 110}, 1 (1984); H. Haber and G. L. Kane, {\it ibid.}
{\bf 117}, 75 (1985).
\bibitem{gutyuka} H. Arason {\it et al.}, Phys. Rev. Lett. {\bf 67}, 2933 (1991); A. Giveon, L. J.
Hall, and U. Sarid, Phys. Lett. B {\bf 271}, 138 (1991).
\bibitem{topcmod} W. A. Bardeen, C. T. Hill, and M. Lindner, Phys. Rev. D {\bf 41}, 1647 (1990).
\bibitem{modelIII} T. Cheng and M. Sher, Phys. Rev. D {\bf 35}, 3484 (1987); M. Sher and Y. Yuan, {\it
ibid.} {\bf 44}, 1461 (1991); J. L. D\'{\i}az Cruz and G.~Lopez~Castro, Phys. Lett. B {\bf 301}, 405
(1993).
\bibitem{ourfcnc} J. L. D\'{\i}az Cruz {\it et al.}, Phys. Rev. D {\bf 41}, 891 (1990).
\bibitem{tcV} See for instance: G. Eilam, J. L. Hewett, and A. Soni, Phys. Rev. D {\bf 44}, 1473
(1991).
\bibitem{PDG98} Particle Data Group, C. Caso {\it et al.}, Eur. Phys. J. C {\bf 3}, 1 (1998).
\bibitem{tbWZinSM} R. Decker, M. Nowakowski,and A. Pilaftsis, Z. Phys. C {\bf 57}, 339 (1993); G.
Mahlon and S. Parke, Phys. Lett. B {\bf 347}, 394 (1995); E. Jenkins, Phys. Rev. D {\bf 56}, 458
(1997); G. Mahlon, ``Theoretical Expectations in Radiative Top Decays'', talk given at the Physics at
Run II: Workshop on Top Physics, Batavia, IL, (1998).
\bibitem{phenoextSM} T. G. Rizzo, Phys. Rev. D {\bf 41}, 1504 (1990); K. Cheung, R. J. N. Phillips,
and A. Pilaftsis, {\it ibid.} {\bf 51}, 4731 (1995); D.~A.~L\'opez~Falc\'on and J. L. D\'{\i}az Cruz,
AIP Conf. Proc. {\bf 490}, 374 (1999).
\bibitem{GVW} For a detailed discussion of models with Higgs triplets, see J. F. Gunion, R. Vega, and
J. Wudka , Phys. Rev. D {\bf 42}, 1673 (1990).
\bibitem{HiggsHG90} J. F. Gunion, H. E. Haber, G. Kane, and S. Dawson. {\it The Higgs Hunter's Guide},
(Addison--Wesley, Reading, MA, 1990), Chap 4, p. 203.
\bibitem{modelIIIpheno} J. L. D\'{\i}az Cruz, J. J. Godina, and G. L\'opez Castro, Phys. Rev. D {\bf
51}, 5263 (1995); D. Atwood, L. Reina, and A. Soni, {\it ibid.} {\bf 55}, 3156 (1997).
\bibitem{ourtcvv} J. L. D\'{\i}az Cruz {\it et al.}, Phys. Rev. D {\bf 60}, 11 5014 (1999).
\bibitem{rhostriction} T. G. Rizzo, Mod. Phys. Lett. A {\bf 6}, 1961 (1991).

\end{references}
\end{document}